\documentstyle[12pt]{article}

\begin{document}

\begin{center}
{\LARGE{\bf A THEORETICAL APPROACH TO

\vspace{1cm}

THE $(\pi^{-}, \gamma \gamma )$
REACTION IN

\vspace{1cm}

NUCLEI}}
\end{center}

\vspace{2cm}

\begin{center}
{\large{ A. Gil and E. Oset}}
\end{center}

\begin{center}
{\small{\it Departamento de F\'{\i}sica Te\'orica and IFIC\\
Centro Mixto Universidad de Valencia - CSIC\\
46100 Burjassot (Valencia) Spain}}
\end{center}

\vskip 1cm
\begin{abstract}
We have studied the $\pi^{-}$ capture in nuclei
leading to two photons, using improved many body methods which have been
tested with success in one photon $\pi^{-}$ radiative capture, $\mu^{-}$
capture and $\nu$ scattering. The qualitative features of the experimental
data are reproduced but there are still some disagreement at small
relative photon angles. The reaction is a potential source of information
on details of the $\gamma \gamma \pi \pi$ vertex where chiral
corrections can be  relevant. New and more precise experiments are planned
which make calculations like the present one relevant and opportune.
\end{abstract}

\newpage

The $(\pi^{-}, \gamma \gamma )$ reaction in nuclei was the subject of much
attention in the last decade \cite{1,2,3,4}. One of the things which
stimulated
this research was the possibility of finding precritical phenomena \cite{5,6},
tied to pion condensation \cite{7,8}, through the nuclear renormalization of
the virtual pions with small energy and a finite momenta which appear in the
driving mechanism for the process \cite{2,4}. The experimental work on this
reaction has been sparse, with only two devoted experiments which provide the
angular correlations of this pionic capture mode from pionic atoms \cite{9}
\cite{10}.
The agreement of the qualitative results of \cite{1,2} with experiment was
only rough, with discrepancies of the order of a factor two to four, and a
poor reproduction of the angular dependence. A more quantitative approach was
followed in ref. \cite{11} using pionic wave functions appropriate for finite
nuclei. Yet the approach relies upon the closure sum, although some corrections
to improve it have also been done, and uses approximate pionic wave functions
which rely upon the concept of $Z_{eff}$ and distortion factors tested in
$\mu^{-}$ capture or radiative pion capture. The average nuclear excitation
energy is also taken as suited for $\mu^{-}$ capture, but in this case the
nuclear excitation energy is smaller since the two photons carry most of the
energy of the pion. The results of \cite{11} agree with experiment at small
relative angles between the photons but disagree in about a factor four at
large angles and the shape of the angular distribution is poorly reproduced.

With time passing there are general reasons to look back to this reaction: the
renormalization of virtual pions in the medium is an important issue to
complement our knowledge acquired in the study of real pions in the meson
factories. This knowledge is important to evaluate the renormalization of weak
currents in nuclei among others. These renormalizations are important to get
proper rates in reaction like $\mu^{-}$ capture or neutrino scattering on
nuclei. With the advent of new neutrino detectors, precise evaluations of
neutrino nucleus cross sections are necessary to calibrate such detectors and
the cross sections are rather sensitive to these nuclear renormalizations
\cite{12}. From the theoretical side, chiral perturbation theory provides
corrections from hadronic loops which can be tested experimentally. Concretely
in the $\gamma \gamma \rightarrow \pi \pi$ process, which enters the driving
term of our reaction, as we shall see, these corrections have been done
\cite{13}, although in a different channel than the one occurring here, and
compared to recent DAPHNE measurements \cite{14}.

Improved techniques allow now to make more complete and precise measurements
of the $(\pi^{-}, \gamma \gamma )$ reaction and new  proposals are under
way \cite{15}.
On the other hand in the last years theoretical progress has been done
which allows a more accurate evaluation of the capture rates than it was
possible in the past.

We follow here a procedure which has proved very accurate and simple to
evaluate radiative pion capture \cite{16}, $\mu^{-}$ capture
 \cite{17}  and
$\nu$ scattering on nuclei \cite{18}. The method consists in evaluating
(let us take $\mu^{-}$ capture as an example) the decay width of a $\mu^{-}$
in infinite nuclear matter but taking into account Pauli blocking, Fermi
motion and the explicit sum over occupied states,
with proper account of the energy of all nucleon states.
Hence the closure sum, and consequently the
dependence of the results on the average excitation energy,
are avoided here. One has also the advantage
of working with the relativistic operators throughout without the need to make
the usual nonrelativistic reduction. At the end the width is evaluated as a
function of the nuclear density and a mapping into finite nuclei is made via
the local density approximation.

For our particular case let us assume a $\pi^{-}$ in dilute nuclear matter
to begin with (proper corrections will be implemented later).
The pion decay width into the $2\gamma$ emission channel is given by

\begin{equation}
\Gamma = \sigma v_{rel} \rho_{p}
\end{equation}

\noindent
with $\rho_{p}$ the proton density and $\sigma$ the $\pi^{-} p \rightarrow
\gamma \gamma n$ cross section. We take the model of fig. 1 for the amplitude
of this latter reaction. Other terms with radiation from the nucleons are much
smaller than these \cite{4,11}. In addition we shall work in the Coulomb
gauge, $\epsilon^{0} = 0, \quad \vec{\epsilon} (k) \vec{k} = 0 \,$, and, since
the momenta of the pionic atoms is small, the terms of fig. 1b, 1c become
negligible. They are exactly zero when $\vec{q} = 0$, and for
$\vec{q} \neq 0$ they are of
the order of $ |\vec{q}/\mu|^{2}$, with $\mu$ the pion mass, which is very
small for
pionic atoms. (Do not confuse these terms with $\nabla^{2}$ terms in \cite{11}
which come from the way the closure sum is done). The expression for
$\sigma v_{rel}$, with $M / E \simeq 1$ for nucleons, is

\begin{eqnarray}
\sigma v_{rel}& =& \int \frac{d^{3}k_{1}}{(2\pi )^{3}} \int \frac{d^{3} k_{2}}
{(2\pi )^{3}} \frac{1}{2k_{1}} \frac{1}{2k_{2}} \frac{1}{2\omega_{\pi}}
\frac{1}{2} \sum^{-} \sum |T|^{2}. \nonumber \\
&&2 \pi \delta (q^{0} + E_{p} - E_{n} - k_{1} - k_{2})
\end{eqnarray}

\noindent
where

\begin{equation}
-i T = 2 i e^{2} \epsilon^{\mu} (k_{1}) \epsilon_{\mu} (k_{2})
\frac{i}{q'^{2}-\mu^{2}} \frac{f}{\mu} \sqrt{2} \vec{\sigma} \vec{q'}
\end{equation}

\noindent
with $q' = q - k_{1} - k_{2}$ and the factor $\frac{1}{2}$ in front of the sum
and average over polarizations stands because of the symmetry of the two
photons. Note that because we work now in a particular gauge

\begin{equation}
\sum_{i=1,2} \epsilon_{i} (k_{1}) \epsilon_{j} (k_{1}) = \delta_{ij} -
\hat{k}_{i} \hat{k}_{j}
\end{equation}

\noindent
Next step consists in replacing $- \pi \rho_{p} \delta (q^{0} + E_{p} -
E_{n} - k_{1} - k_{2} )$ by $Im \bar{U} (q^{0}- k_{1} - k_{2}, -(\vec{k}_{1} +
\vec{k}_{2}))$, the Lindhard function for ph excitation of the np type given
by

\begin{equation}
\bar{U} (q')= 2 \int \frac{d^{3}p}{(2 \pi )^{3}}
\frac{n_{1} (\vec{p}\:) [1-n_{2} (\vec{p} + \vec{q}\,' )]}
{q^{0} + E_{p} (\vec{p}\:) - E_{n} (\vec{p} + \vec{q}\, ') + i \epsilon}
\end{equation}

\noindent
with $n_{1} (\vec{p}\:)$ the occupation number for protons and $n_{2} (\vec{p}
+
\vec{q}\, ')$ the occupation number for neutrons.
This substitution takes into account the finite density corrections,
accounting explicitly for Fermi motion and Pauli blocking.
After the proper substitutions
we obtain
\begin{eqnarray}
\frac{d \Gamma}{d \Omega_{12}}&=& \frac{1}{\mu} \frac{1}{(2 \pi )^{5}}
\left(\frac{f}{\mu}\right)^{2}
e^{4} \int k_{1} d k_{1} \int k_{2} d k_{2} \vec{q'}^{2}
(1 + cos^{2} \theta_{12}).\nonumber \\
&&\Bigl(\frac{1}{q'^{2} - \mu^{2}}\Bigr)^{2} (-2)
Im \bar{U} (q', \rho_{p}, \rho_{n})
\end{eqnarray}

\noindent
with $\theta_{12}, \Omega_{12}$, the relative angle and solid angle of the two
photons. In the actual evaluation of eq.(6) we consider a lower threshold for
$k_{1}$ and $k_{2}$ of 25 MeV in order to compare with the data of the
experiment \cite{9}. For $\pi^{-}$ capture from a particular pionic orbit
we have

\begin{equation}
\Gamma_{nl} = \int d^{3} r | \phi_{nl} (\vec{r}\:)|^{2} \Gamma (\rho_{p}
(\vec{r}\:) \rho_{n} (\vec{r}\:))
\end{equation}

\noindent
which makes explicit use of the local density approximation.
$\phi_{nl}(\vec{r}\:)$ are the pionic wave functions which we obtain by solving
the Klein Gordon equation with the potential of ref. \cite{19}.

Finally, since the experiment does not distinguish the decay from individual
pionic orbits, a weighed average like in radiative $\pi^{-}$ capture must
be done and we get

\begin{equation}
R^{\gamma \gamma} = \sum_{nl} \frac{\Gamma^{\gamma \gamma}_{nl}}{\Gamma_{nl}
^{abs}} \omega_{nl}
\end{equation}

\noindent
We take $\omega_{nl}$ from \cite{20}
(see also \cite{16})
and $\Gamma_{nl} ^{abs}$ from several
experiments tabulated in \cite{17}.

We can also take into account the renormalization of the process due to medium
effects in the pion propagator, in analogy to the propagation of $ ph$
components
in the longitudinal channel. The details can be seen in ref. \cite{17} and it
amounts to the change

\begin{equation}
Im \bar{U} \rightarrow \frac{Im \bar{U}}{|1 - U V_{l}|^{2}}
\end{equation}

\noindent
with $U$ the Lindhard function for
$ph$ plus $\triangle h$ excitation (and
different normalization than $\bar{U}$, an extra factor 2
in the ph excitation part
 to account for isospin)
and $V_{l}$ the longitudinal part of the $ph$ interaction. One must be cautious
here. Since the two photons will carry most of the pion energy, the energy
left for nuclear ph excitation is small.
Then if the photons go back to back, the momentum transfer to the ph
excitation is also small. This situation, with $q^0$,$\vec{q}$ small leads
to  unrealistic values of ReU$_{N}$, from ph excitation, if standard
formulas \cite{21} are used. Indeed, for q$^0$=0 and $\vec{q}\rightarrow
0$,  ReU$_{N}$ has a finite limit which is fallacious since the
response in finite nuclei is strictly zero in closed shell nuclei.
The discrepancies appear because one has a ratio of a numerator
which is zero and a denominator which contains the ph excitation
energy. The latter one is finite in finite nuclei, but zero in the
continuum spectrum of infinite matter. The problem is solved if
a realistic excitation gap energy is considered in the evaluation
of ReU$_{N}$ and this is done in \cite{22}. We use for ReU$_{N}$ the
expressions of this latter reference.

Our results, compared with the experimental ones of ref. \cite{9} are shown in
fig. 2 for  $^{9}Be$ and fig. 3 for $^{12}C$.
Our results approximately agree with experiment for angles
above 90$^0$. In the case of $^9$Be the results are on the upper side of
the data, while for $^{12}C$ they are on the lower side. At small angles,
however, our results are consistently below the data. The shape of the
angular distribution is qualitatively correct, something that did not
appear in the previous theoretical calculations \cite{1,2,11}.
The renormalization of the longitudinal response, eq. (9), in this case
reduces the results, particularly at large angles, leading to
a better agreement with the data.

	We hope, however, that the improved techniques in the new proposals
\cite{15} lead to a new wave of very precise data from where one could take
 more seriously the discrepancies with theoretical models which
would allow us to make progress on details of the elementary
$\gamma \gamma \pi\pi$ vertices or possible missing many body effects.

\vskip 2.5cm

This work has been partially supported by CICYT contract no. AEN 93-1205

One of us A.G. wishes to acknowledge financial support from the
Ge\-ne\-ra\-litat
Valenciana in the program of Formaci\'on de Personal Investigador.

\newpage

\centerline{\Large \bf Figure Captions}
\vskip 1cm
\noindent
{\Large \bf Fig.1} Feynman diagrams for the amplitude.
\vskip 0.7cm
\noindent
{\Large \bf Fig.2} Comparison between experimental data
and our theoretical results
for $^{9}$Be.
 Energy resolution in the experimental data
is 25 MeV
photon threshold.
 Energy resolution is included in the theoretical results for
  25 MeV  (solid line) and 17 MeV (long dashed dotted line)
 photon threshold.
\vskip 0.7cm
\noindent
{\Large \bf Fig.3} Comparison between experimental
data and our theoretical results
for $^{12}$C.
 Energy resolution in the experimental data
is 25 MeV (boxes) and 17 MeV (crosses)
photon threshold.
Energy resolution is included in the theoretical results for
  25 MeV  (solid line) and 17 MeV (long dashed dotted line)
 photon threshold.

\end{document}